\documentclass[12pt]{article} 
\usepackage{graphicx}
\begin{document}
\baselineskip 10mm

\centerline{\bf Rabi oscillations in the four-level double-dot
structure}\centerline{\bf under the influence of the resonant
pulse}

\vskip 2mm

\centerline{Alexander V. Tsukanov}

\vskip 2mm

\centerline{\it Institute of Physics and Technology, Russian
Academy of Sciences}
\centerline{\it Nakhimovsky pr. 34, Moscow
117218, Russia}
\centerline{\it E-mail: tsukanov@ftian.oivta.ru}

\vskip 4mm

\begin{abstract}

We study theoretically the quantum dynamics of an electron in the
symmetric four-level double-dot structure under the influence of
the monochromatic resonant pulse. The probability amplitudes of
the eigenstates relevant for the quantum dynamics are found from
the solution of the non-stationary Schr\"odinger equation. The
first-order correction term to the solution obtained through the
rotating wave approximation is calculated. The three-level
double-dot dynamics and the two-level single-dot dynamics, as well
as the off-resonant excitation process, are derived from the
general formulae for corresponding choices of the pulse and
structure parameters. The results obtained may be applied to the
solid-state qubit design.

\end{abstract}

\vskip 4mm

PACS number(s): 03.67.Lx, 73.21.La, 73.23.-b, 78.67.Hc

\vskip 6mm

\newpage

\centerline{\bf I. INTRODUCTION}

  Recently the low-dimensional semiconductor structures
containing a small number of electrons in the size-quantized
conduction band have attracted much attention. One of the reasons
for that interest consists in their potential applicability to the
quantum information processing \cite{1}. It is commonly believed
that those structures may be scaled up to form the quantum
register with appropriate number of qubits. During the past decade
many proposals for the semiconductor qubit realization were made.
Here we mention the phosphorous donors embedded in a silicon host
\cite{2,3} and a wide class of the systems based on the quantum
dots (QDs) (see, e.g., \cite{4,5,6,7,8,9,10,11,12,13,14}).

One of the features characterizing the coherent evolution of a
quantum system is the Rabi oscillations induced by an external
field. If the system was initially in one of the eigenstates of
unperturbed Hamiltonian, it starts to oscillate under perturbation
between its eigenstates giving rise to a broad class of different
phenomena observed in the experiments. In view of the quantum
algorithm realization, the Rabi oscillations are often considered
as the process generating the desired qubit-state evolution. The
most popular setup to study the Rabi oscillations of the electron
population in the semiconductor nanostructures is based on the
double-dot structure containing a single electron in the
conduction band \cite{15,16} or, alternatively, on the artificial
$H_{2}^{+}$ molecule formed by the donor implantation techniques
(the so-called "charge qubit" \cite{17}). In what follows we shall
consider the QDs keeping in mind the general character of the
conclusions made below. The standard method usually exploited for
an electron charge manipulations in such structures is based on
the electrostatic field control.  By applying an adiabatically
switched voltage one can lower or raise the potential barrier that
separates the QDs thus allowing an electron to tunnel  between the
QDs or localizing it in one of the QDs. The quantum evolution of
an electron clearly demonstrates the Rabi-like behavior that
originates from the coherent electron tunnelling between the QDs
\cite{15,16}.

There is another scheme for the manipulations with a single
electron confined in the double-dot structure. It is based on the
resonant interaction of an electron with the coherent
electromagnetic pulse \cite{18}. The laser field, instead of
electrostatic one, is applied to the double-dot structure, and the
 quantum state engineering is realized via the optically-induced
transitions between the size-quantized electron levels. As it was
shown, the pulse parameters (the frequency, duration, and
amplitude) can be chosen so as to drive the electron, localized
initially in the ground state in one of the QDs, to the ground
state of another QD via the state delocalized over QDs and used
here as a "transport" state. If the states localized in different
QDs are viewed as the Boolean states 0 and 1, then, e. g., the
electron transfer between them may be considered as the unitary
operation NOT. The idea was initially proposed by Openov \cite{18}
and then developed further in the works \cite{19,20,21,22,23,24}.
The influence of strong electromagnetic fields on the tunnelling
phenomena in several-level nanostructures was studied in Refs.
\cite{25,26}. It was shown that a laser with appropriate power and
frequency can drive the electron between quantum wells in a
finite-size quantum well nanostructure or localize it in one of
them. In Ref. \cite{18} the opposite effect - i.e., a
weak-laser-induced electron transfer between two quantum dots, was
considered.
 In that model of the one-electron quantum dynamics, the assumption
of instantaneous spreading of an excited electron over the QD
structure was made or, equivalently, the matrix element of optical
dipole transition between the ground and the "transport" states
was supposed to be much less than the matrix element for
tunnelling of an excited electron between the QDs. This allowed to
describe the dynamics within the framework known in the atomic
optics as the three-level (or $\Lambda$) scheme. The probability
amplitudes to find an electron in the states localized in each of
the QDs and in the "transport" state delocalized over the QD
structure, were found as functions of the time, the pulse
parameters, and the structure parameters. Provided that the
quantum evolution is coherent, this process describes the
three-level Rabi oscillations of the electron population. However,
in all of these studies,  the rigorous quantitative analysis of
the assumption concerning the instantaneous spreading of an
excited electron over the double-dot structure has not been
presented.

In this work we focus on the quantitative study of the coherent
quantum dynamics of an electron in the symmetric four-level
double-dot
 structure under the influence of the resonant laser
pulse for arbitrary tunnelling rates between the excited states of
the QDs. Here we give the detailed derivation of the analytical
expressions for the probability amplitudes of the electron
eigenstates relevant for the quantum dynamics. The results will be
presented in terms of the basis states of isolated QDs. In this
picture, the probability amplitudes are the explicit functions of
both the tunnelling matrix element and the matrix element of the
electron optical dipole transition. We show that the character of
the system evolution is determined by the ratio between these
matrix elements. The three-level double-dot and the two-level
single-dot dynamics are derived from the general formulae as the
limiting cases. We examine also the off-resonant excitation scheme
that is very promising for the experimental realization of the
proposed method of electron state manipulation.

The paper is organized as follows. In Section II we present the
description of the model and obtain the general solution for the
electron dynamics in the four-level double-dot structure. The
important particular cases of the three- and two-level dynamics as
well as the off-resonant Raman-like excitation are considered in
Section III. Section IV contains the results of numerical
simulations. The conclusions are summarized in Section V.

\vskip 5mm

\centerline{\bf II. MODEL AND GENERAL SOLUTION}

Let us consider the double-dot structure (see Fig.1)
 containing an electron in the size-quantized conduction band. For
the sake of simplicity, we suppose the dots $A$ and $B$ to be
identical. The existence of at least two one-electron orbital
states $\left| {A\left( B \right)0} \right\rangle $ and $\left|
{A\left( B \right)1} \right\rangle $ (ground and excited) in each
of the QDs is assumed, with the one-electron wave functions
$\varphi _{A(B)0} \left( {\bf{r}} \right)=\left\langle {{\bf{r}}}
 \mathrel{\left | {\vphantom {{\bf{r}} {A(B)0 }}}
 \right. \kern-\nulldelimiterspace}
 {{A(B)0 }} \right\rangle
$ and $\varphi _{A(B)1} \left( {\bf{r}}\right )=\left\langle
{{\bf{r}}}
 \mathrel{\left | {\vphantom {{\bf{r}} {A(B)1 }}}
 \right. \kern-\nulldelimiterspace}
 {{A(B)1 }} \right\rangle $, respectively.
Provided that the distance between the QDs is sufficiently large,
the wave functions of the QD ground states are localized in
corresponding QDs, and their overlap can be neglected. The overlap
between the ground state and the excited state belonging to
different QDs will be neglected as well: $\left\langle {{A(B)0 }}
 \mathrel{\left | {\vphantom {{A(B)0 } {B(A)1 }}}
 \right. \kern-\nulldelimiterspace}
 {{B(A)1 }} \right\rangle \approx 0$. The excited levels are
chosen to be close to the edge of the potential barrier separating
the QDs. They couple through the electron tunnelling \cite{18}.

 The Hamiltonian of an electron confined in the symmetric four-level double-dot
 structure is
\begin{equation}
\begin{array}{l}
 H_{0} = \varepsilon _0 \left( {\left| {A0} \right\rangle \left\langle {A0} \right| + \left| {B0} \right\rangle \left\langle {B0} \right|} \right) + \varepsilon _1 \left( {\left| {A1} \right\rangle \left\langle {A1} \right| + \left| {B1} \right\rangle \left\langle {B1} \right|} \right) +  \\
 \,\,\,\,\,\,\,\,\, + \left[ { - V \left( t \right)\left| {A1} \right\rangle \left\langle {B1} \right| +  H.c.} \right], \\
 \end{array}
\end{equation}
where $\varepsilon _0  $ and $\varepsilon _1  $ are the
one-electron energies of the ground and excited states,
respectively (the same for both QDs); $V \left( t \right)>0$
 is the matrix element for the electron tunnelling between the excited
states of the QDs, that, in general, may be a time-dependent
function.

We consider the quantum evolution of an electron under the
influence of the electromagnetic field that induces the optical
transitions between the ground and excited states in each of the
QDs ($\left| {A0} \right\rangle  \leftrightarrow \left| {A1}
\right\rangle $ and $\left| {B0} \right\rangle  \leftrightarrow
\left| {B1} \right\rangle $). It is convenient to examine that
evolution as a complex process including both the optical
excitation of an electron in one of the QDs and the tunnelling of
the excited electron into the other QD.
 The model Hamiltonian has the form
\begin{equation}
 H = H_0  +  \left[ {\lambda \left( t \right)\left( {\left| {A0} \right\rangle \left\langle {A1} \right| + \left| {B0} \right\rangle \left\langle {B1} \right|} \right)+  H.c.}
 \right],
 \end{equation}
where $\lambda \left( t \right)$ is the matrix element of the
electron-field interaction. In what follows we shall suppose the
matrix elements $V$ and $\lambda$ to be real and not show
explicitly that they are time-dependent. The criterion of
applicability of the model Hamiltonian (2) is expressed by the
inequalities $\left| \lambda \right| \ll \omega _{10} $, $V \ll
\omega _{10} $, where $\omega _{10} = \varepsilon _1 - \varepsilon
_0 $. Besides, we shall suppose that there are no additional
levels localized in the near neighborhood of the energy
$\varepsilon _1$.

 The state vector of the system may be represented in  terms of the
eigenstates of isolated QDs as
\begin{equation}
 \left| {\Psi \left( t \right)} \right\rangle  = \sum\limits_{n =
A0,B0,A1,B1} {c_n \left( t \right)\left| n \right\rangle }.
\end{equation}
 Let an electron be initially localized in the ground state of the QD A, i.e. $\,\left| {\Psi \left( 0 \right)} \right\rangle  = \left| {A0} \right\rangle $.
The quantum evolution of the state vector is governed by the
non-stationary Schr\"odinger equation
\begin{equation}
 i\frac{{\partial \left| {\Psi \left( t \right)} \right\rangle }}{{\partial t}} = H\left| {\Psi \left( t \right)}
 \right\rangle,
 \end{equation}
 or, in the matrix form,
\begin{equation}
i\frac{\partial }{{\partial t}}\left( {\begin{array}{*{20}c}
   {c_{A0} }  \\
   {c_{B0} }  \\
   {c_{A1} }  \\
   {c_{B1} }  \\
\end{array}} \right) = \left( {\begin{array}{*{20}c}
   {\varepsilon _0 } & 0 & \lambda  & 0  \\
   0 & {\varepsilon _0 } & 0 & \lambda   \\
   \lambda  & 0 & {\varepsilon _1 } & { - V}  \\
   0 & \lambda  & { - V} & {\varepsilon _1 }  \\
\end{array}} \right)\left( {\begin{array}{*{20}c}
   {c_{A0} }  \\
   {c_{B0} }  \\
   {c_{A1} }  \\
   {c_{B1} }  \\
\end{array}} \right)
\end{equation}
 with the initial conditions $c_n \left( 0 \right) = \delta _{n,\,A0} $
(hereafter $\hbar  = 1$).

The straightforward diagonalization of the Hamiltonian matrix
amounts to the set of eigenvectors
\begin{equation}
\begin{array}{l}
 \left| 1 \right\rangle  = u_ -  {{\left( {\left| {A0} \right\rangle  + \left| {B0} \right\rangle } \right)} \mathord{\left/
 {\vphantom {{\left( {\left| {A0} \right\rangle  + \left| {B0} \right\rangle } \right)} {\sqrt 2 }}} \right.
 \kern-\nulldelimiterspace} {\sqrt 2 }} + v_ -  {{\left( {\left| {A1} \right\rangle  + \left| {B1} \right\rangle } \right)} \mathord{\left/
 {\vphantom {{\left( {\left| {A1} \right\rangle  + \left| {B1} \right\rangle } \right)} {\sqrt 2 }}} \right.
 \kern-\nulldelimiterspace} {\sqrt 2 }}, \\
 \left| 2 \right\rangle  = u_ +  {{\left( {\left| {A0} \right\rangle  - \left| {B0} \right\rangle } \right)} \mathord{\left/
 {\vphantom {{\left( {\left| {A0} \right\rangle  - \left| {B0} \right\rangle } \right)} {\sqrt 2 }}} \right.
 \kern-\nulldelimiterspace} {\sqrt 2 }} + v_ +  {{\left( {\left| {A1} \right\rangle  - \left| {B1} \right\rangle } \right)} \mathord{\left/
 {\vphantom {{\left( {\left| {A1} \right\rangle  - \left| {B1} \right\rangle } \right)} {\sqrt 2 }}} \right.
 \kern-\nulldelimiterspace} {\sqrt 2 }}, \\
 \left| 3 \right\rangle  = v_ -  {{\left( {\left| {A0} \right\rangle  + \left| {B0} \right\rangle } \right)} \mathord{\left/
 {\vphantom {{\left( {\left| {A0} \right\rangle  + \left| {B0} \right\rangle } \right)} {\sqrt 2 }}} \right.
 \kern-\nulldelimiterspace} {\sqrt 2 }} - u_ -  {{\left( {\left| {A1} \right\rangle  + \left| {B1} \right\rangle } \right)} \mathord{\left/
 {\vphantom {{\left( {\left| {A1} \right\rangle  + \left| {B1} \right\rangle } \right)} {\sqrt 2 }}} \right.
 \kern-\nulldelimiterspace} {\sqrt 2 }}, \\
 \left| 4 \right\rangle  = v_ +  {{\left( {\left| {A0} \right\rangle  - \left| {B0} \right\rangle } \right)} \mathord{\left/
 {\vphantom {{\left( {\left| {A0} \right\rangle  - \left| {B0} \right\rangle } \right)} {\sqrt 2 }}} \right.
 \kern-\nulldelimiterspace} {\sqrt 2 }} - u_ +  {{\left( {\left| {A1} \right\rangle  - \left| {B1} \right\rangle } \right)} \mathord{\left/
 {\vphantom {{\left( {\left| {A1} \right\rangle  - \left| {B1} \right\rangle } \right)} {\sqrt 2 }}} \right.
 \kern-\nulldelimiterspace} {\sqrt 2 }}, \\
 \end{array}
\end{equation}
and to the corresponding eigenenergies
\begin{equation}
\begin{array}{l}
 E_1  = \varepsilon _0  + \frac{1}{2}\left( {\omega _ -   + \sqrt {\omega _ - ^2  + 4\lambda ^2 } } \right),\,\,E_2  = \varepsilon _0  + \frac{1}{2}\left( {\omega _ +   + \sqrt {\omega _ + ^2  + 4\lambda ^2 } } \right), \\
 E_3  = \varepsilon _0  + \frac{1}{2}\left( {\omega _ -   - \sqrt {\omega _ - ^2  + 4\lambda ^2 } } \right),\,\,E_4  = \varepsilon _0  + \frac{1}{2}\left( {\omega _ +   - \sqrt {\omega _ + ^2  + 4\lambda ^2 } } \right), \\
 \,\,\,\,\,\,\,\,\,\,\,\,\,\,\,\,\,\,\,\,\,\,\,\,\,\,\,\,\,\,\,\,\,\,\,\,\,\,\,\,\,\,\,\,\,\,\,\,\,\,\,\,\,\,\,\,\,\,\ E_4<E_3<...,
 \end{array}
\end{equation}
where $u_ \pm \approx \frac{\lambda }{{\omega _ \pm  }}\left( {1 -
\frac{3}{2}\frac{{\lambda ^2 }}{{\omega _ \pm ^2 }}}
\right),\,\,v_ \pm   \approx 1 - \frac{1}{2}\frac{{\lambda ^2
}}{{\omega _ \pm ^2 }}$ and $\omega _ \pm   = \omega _{10}  \pm
V$. (Here we restrict ourselves by the third-order terms in the
parameters $\frac{\left| \lambda \right|}{{\omega _ \pm  }}\ll 1$.
Note that the expansion of (7) over the small parameters
$\frac{\left| \lambda \right|}{{\omega _ \pm  }}\ll 1$ gives rise
to the Bloch-Siegert term $\frac{{\lambda ^2 }}{{\omega _ {10} }}$
in the eigenenergies $E_i , i=1-4$.) The matrix diagonalizing the
right-hand side of Eq. (5) has the form
\begin{equation}
D = \left( {\begin{array}{*{20}c}
   U_1 & U_2  \\
   U_2 & { - U_1}  \\
\end{array}} \right),
\end{equation}
where
\begin{equation}
U_1 = \frac{1}{\sqrt 2}\left( {\begin{array}{*{20}c}
   {u_ -  } & {u_ +  }  \\
   {u_ -  } & { - u_ +  }  \\
\end{array}} \right),\,\,U_2 = \frac{1}{\sqrt 2}\left( {\begin{array}{*{20}c}
   {v_ -  } & {v_ +  }  \\
   {v_ -  } & { - v_ +  }  \\
\end{array}} \right)
\end{equation}
(The columns of $D$ are the transposed eigenvectors $\left| { k }
\right\rangle , k=1-4$).

The electron dynamics is most easily revealed in the
representation of the instantaneous eigenstates, Eqs. (6), of the
system Hamiltonian. In this basis, the time-dependent state vector
of the system takes the form
\begin{equation}
\left| \Phi\left( t \right)  \right\rangle  = \sum\limits_{k =
1}^4 {a_k \left( t \right)\left| { k } \right\rangle },
\end{equation}
where the instantaneous probability amplitudes ${a_k \left( t
\right)},\, k=1-4$, are related to the probability amplitudes
${c_n \left( t \right)}$ by the matrix $D$:
\begin{equation}
{\bf{c}}\left( t \right) = D{\bf{a}}\left( t \right),
\end{equation}
where ${\bf{c}}\left( t \right) = \left[ {c_{A0} \left( t
\right),c_{B0} \left( t \right),c_{A1} \left( t \right),c_{B1}
\left( t \right)} \right]^T$ and ${\bf{a}}\left( t \right) =
\left[ {a_1 \left( t \right),a_2 \left( t \right),a_3 \left( t
\right),a_4 \left( t \right)} \right]^T $.

The Schr\"odinger equation in the new basis reads
\begin{equation}
i\frac{{\partial \left| \Phi \left( t \right)  \right\rangle
}}{{\partial t}} = \left\{ {D^ \dagger  HD - iD^ \dagger
\frac{{\partial D}}{{\partial t}}} \right\}\left| \Phi \left( t
\right) \right\rangle
\end{equation}
or, in the matrix form,
\begin{equation}
i\frac{\partial }{{\partial t}}\left( {\begin{array}{*{20}c}
   {a_1 }  \\
   {a_2 }  \\
   {a_3 }  \\
   {a_4 }  \\
\end{array}} \right) = \left( {\begin{array}{*{20}c}
   {E_1 } & 0 & {\mu _ -  } & 0  \\
   0 & {E_2 } & 0 & {\mu _ +  }  \\
   {\mu _ -  ^* } & 0 & {E_3 } & 0  \\
   0 & {\mu _ +  ^* } & 0 & {E_4 }  \\
\end{array}} \right)\left( {\begin{array}{*{20}c}
   {a_1 }  \\
   {a_2 }  \\
   {a_3 }  \\
   {a_4 }  \\
\end{array}} \right),
\end{equation}
with the initial conditions $a_{1\left( 2 \right)} \left( 0
\right) = u_{-\left( + \right)}/{\sqrt 2} ,\,\,\,a_{3\left( 4
\right)} \left( 0 \right) = v_{-\left( + \right)}/{\sqrt 2}$. Here
$E_{1\left( 2 \right)}  \approx \varepsilon _0  + \omega _{ -
\left(  +  \right)} \left( {1 + \frac{{\lambda ^2 }}{{\omega _{ -
\left(  +  \right)}^2 }}} \right),\,E_{3\left( 4 \right)}  \approx
\varepsilon _0  - \frac{{\lambda ^2 }}{{\omega _{ - \left(  +
\right)} }}$ and $\mu _ \pm \approx i\frac{\partial }{{\partial
t}}\left[ {\left( {\frac{\lambda }{{\omega _ \pm  }}}
\right)\left( {1 - \frac{{4\lambda ^2 }}{{\omega _ \pm ^2 }}}
\right)} \right]$, in agreement with the approximations for
$u_{\pm}$ and $v_{\pm}$. Besides, we have neglected the terms $
\sim \frac{{\lambda ^3 }}{{\omega _ \pm ^3 }}\frac{\partial
}{{\partial t}}\left( {\frac{\lambda }{{\omega _ \pm  }}} \right)$
in $E_i$ assuming them much smaller than the Bloch-Siegert term.
This approximation requires that $\left| {\frac{\lambda }{{\omega
_ \pm ^2 }}\frac{\partial }{{\partial t}}\left( {\frac{\lambda
}{{\omega _ \pm  }}} \right)} \right| \ll 1$ and imposes the
constrain on the speed of changes in both the tunnelling matrix
element $V$ and the matrix element of the electron-field
interaction $\lambda$. The terms $ \sim \frac{{\lambda ^4
}}{{\omega _ \pm ^4 }}\frac{\partial }{{\partial t}}\left(
{\frac{\lambda }{{\omega _ \pm  }}} \right)$ in $\mu _ \pm$ have
been rejected as well.
 At this stage of consideration we
see that the pairs of coefficients ($a_1$, $a_3$) and ($a_2$,
$a_4$) evolve independently from each other. The four-level
problem of Eq. (5) thus reduces to the two-level ones that can, in
principle, be solved by one of the standard methods developed
earlier \cite{27}.

To proceed further we have to concretize the electron-field
interaction term $\lambda$ in Eq. (2). In what follows we shall
consider the situation where the tunnelling matrix element $V$ is
time-independent. It is worth to mention that for the
adiabatically switched voltages the Hamiltonian matrix in Eq. (13)
is diagonal and its solution is straightforward. As an example we
give here the solution describing the electron behavior under the
influence of the electrostatic field ${\bf{E}}_0\ne {\bf{E}}_0
\left( t \right)$ ($\mu _ \pm = 0$). The expression for ${\bf{c}}$
follows directly from Eqs. (11) and (13). The probability
amplitudes for an electron to be in the ground states of the QDs
have the form
\begin{equation}
c_{A0}  \approx e^{ - i\left( {\varepsilon _0  - \frac{{\lambda_0
^2 }}{{\omega _{10} }}} \right)t} \cos \left( {\frac{{\lambda_0 ^2
V}}{{\omega _{10}^2 }}t} \right),\,\,\,c_{B0}  \approx ie^{ -
i\left( {\varepsilon _0  - \frac{{\lambda_0 ^2 }}{{\omega _{10}
}}} \right)t} \sin \left( {\frac{{\lambda_0 ^2 V}}{{\omega _{10}^2
}}t} \right),
\end{equation}
where $\lambda _0  = e{\bf{E}}_0 \int {{\varphi _{A\left( B
\right)0}^* \left( {\bf{r}} \right){\bf{r}} \varphi _{A\left( B
\right)1} \left( {\bf{r}} \right)} d{\bf{r}}}$
 is the matrix element of optical dipole transition between the
states $\left| A(B)0 \right\rangle $ and $\left| A(B)1
\right\rangle $. The population is therefore localized (up to the
${{\lambda_0 ^2 } \mathord{\left/
 {\vphantom {{\lambda_0 ^2 } {\omega _{10}^2 }}} \right.
 \kern-\nulldelimiterspace} {\omega _{10}^2 }} \ll 1$ terms) in the ground-state subspace
and exhibits the two-level Rabi oscillations at the frequency
$\Omega_{0} = {{\lambda_0 ^2 V} \mathord{\left/
 {\vphantom {{\lambda_0 ^2 V} {\omega _{10}^2 }}} \right.
 \kern-\nulldelimiterspace} {\omega _{10}^2 }}$. The probabilities to find an electron
in the states ${\left| {A1} \right\rangle }$ and ${\left| {B1}
\right\rangle }$ are of order of ${{\lambda_0 ^2 } \mathord{\left/
 {\vphantom {{\lambda_0 ^2 } {\omega _{10}^2 }}} \right.
 \kern-\nulldelimiterspace} {\omega _{10}^2 }} \ll 1$ and oscillate at the frequency $V$.
One sees that even for the static fields, a substantial electron
state evolution in the double-dot structure occurs for the
characteristic time $T \sim 1/ \Omega_{0}$. In principle, the
electron oscillations may be utilized for the qubit-state
engineering but this process seems to be too slow in comparison
with the resonant optical driving \cite{18} and unviable in view
of the decoherence. However, this effect should be taken into
account if quantum operations are performed through the sequence
of electrostatic voltages \cite{9,11} since in this case the
electron transitions between the localized and delocalized states
induced by the static fields bring about an unwanted qubit
dynamics, i.e., a computational error.

The central part of our investigation will be devoted to the
interaction of an electron with the time-dependent resonant
pulses. For the sake of simplicity we consider a monochromatic
square pulse of the amplitude ${\bf{E}}_0$, the duration $T$, and
the frequency $\omega$:
\begin{equation}
{\bf{E}}\left( t \right) = {\bf{E}}_0 \cos \left( {\omega t}
\right)\left[ {\theta \left( t \right) - \theta \left( {t - T}
\right)} \right].
\end{equation}
In this case
\begin{equation}
\lambda = \lambda _0 \cos \left( {\omega t} \right)\left[
{{\rm{\theta }}\left( t \right) - {\rm{\theta }}\left( {t - T}
\right)} \right],
\end{equation}
 where ${{\rm{\theta }}\left( t \right)}$ is the step
function. The frequency $\omega$ of the laser pulse may be detuned
from the resonant frequencies $\omega_\pm$ by the values $\delta _
\pm = \omega  - \omega _ \pm  $, where $\left| {\delta _ \pm }
\right| \ll \omega $.

As we see from Eq. (13), it is sufficient to analyze the dynamics
of just one pair of the coefficients, e.g., $a_1$ and $a_3$ (the
dynamics for $a_2$ and $a_4$  is then revealed by the substitution
$\omega _ - \to \omega _ + $ in the results obtained for $a_1$ and
$a_3$). Transforming the coefficients $a_1$ and $a_3$ according to
the formulae $a_1  = \tilde a_1 e^{ - i\left( {\varepsilon _0  +
\omega _ -  } \right)t}$ and $a_3  = \tilde a_3 e^{ - i\varepsilon
_0 t} $
 and inserting the expression (16) for
$\lambda$ into Eq. (13), we arrive at the set of two coupled
linear differential equations for the coefficients $ \tilde a_1$
and $ \tilde a_3$ (an analogous set is obtained for the
coefficients $ \tilde a_2$ and $ \tilde a_4$):
\begin{equation}
\left\{ \begin{array}{l}
 i\dot {\tilde a_1}  = \frac{{{\lambda '}_0^2 }}{\omega }\cos \left( {\omega t} \right)\tilde a_1  + \frac{{{\lambda '}_0 }}{2}\left( {1 - e^{i2\omega t} } \right)e^{ - i\delta _ -  t} \tilde a_3  \\
 i\dot {\tilde a_3}  =  - \frac{{{\lambda '}_0^2 }}{\omega }\cos \left( {\omega t} \right)\tilde a_3  + \frac{{{\lambda '}_0 }}{2}\left( {1 - e^{ - i2\omega t} } \right)e^{i\delta _ -  t} \tilde a_1  \\
 \end{array} \right.,\,\,\,\,\,\,\,\,\,\,\lambda '_0  = \lambda _0 \frac{\omega }{{\omega _ -  }}.
\end{equation}
Here we restrict ourselves to the terms linear on the small
parameter ${\lambda_0 \mathord{\left/
 {\vphantom {\lambda_0  \omega }} \right.
 \kern-\nulldelimiterspace} \omega }$ in $\mu _ \pm  $ and retain the Bloch-Siegert term $ \sim {{{\lambda '}_0^2 } \mathord{\left/
 {\vphantom {{{\lambda '}_0^2 } \omega }} \right.
 \kern-\nulldelimiterspace} \omega }$
. As we shall see below, the account of this term is necessary for
obtaining the correct result within the first-order approximation
on the parameter ${{\lambda '_0 } \mathord{\left/
 {\vphantom {{\lambda '_0 } \omega }} \right.
 \kern-\nulldelimiterspace} \omega }\ll 1$.

Usually, at this point the rotating wave approximation (RWA) is
made and the fast oscillating exponents ${e^{ \pm i2\omega t} }$
in Eqs. (17) are omitted. The solution thus accounts only of the
one-photon processes conserving the energy of the system. This
approximation is valid if the frequency $\omega _{10} \approx
\omega \approx \omega _ \pm  $ dominates the Rabi frequency that
is of the order of $\left| \lambda _ 0 \right|$. Since this
requirement on the pulse parameters is inherent to many optics
setups, the theoretical predictions based on the RWA are in
excellent agreement with the experimental data. However, any
possible extension of the RWA seems to be very instructive in view
of the quantitative estimate of corrections to the results
obtained by the RWA. In Ref. \cite{28} the authors suggested a
simple and clear way of how to calculate the first-order
correction term to the RWA solution. Using the adiabatic
elimination procedure for the virtual two-photon states, they were
able to find the probability amplitudes for a two-level system
beyond the RWA. Here we shall utilize their method to solve the
set of Eqs. (17).

According to Ref. \cite{28}, the coefficients $ \tilde a_1$ and $
\tilde a_3$ can be sought in the form:
\begin{equation}
\begin{array}{l}
 \tilde a_1  \approx a_1^{\left( 0 \right)}  + a_1^{\left(  -  \right)} e^{ - i2\omega t}  + a_1^{\left(  +  \right)} e^{i2\omega t} \,\,\  , \\
  \\
 \tilde a_3  \approx a_3^{\left( 0 \right)}  + a_3^{\left(  -  \right)} e^{ - i2\omega t}  + a_3^{\left(  +  \right)} e^{i2\omega t} \,\,\  , \\
 \end{array}
\end{equation}
where the higher-order terms proportional to ${e^{ \pm i2m\omega
t} },\,m>1$, are dropped. From the Eqs. (17) and (18) we obtain
the set of six equations:
\begin{equation}
 \left\{ \begin{array}{l}
 i\dot {a_1}^{\left( 0 \right)}  = \frac{{{\lambda '}_0^2 }}{\omega }a_1^{\left( 0 \right)}  + \frac{{{\lambda '}_0 }}{2}e^{ - i\delta _ -  t} \left( {a_3^{\left( 0 \right)}  - a_3^{\left(  -  \right)} } \right) \\
 i\dot {a_3}^{\left( 0 \right)}  =  - \frac{{{\lambda '}_0^2 }}{\omega }a_3^{\left( 0 \right)}  + \frac{{{\lambda '}_0 }}{2}e^{i\delta _ -  t} \left( {a_1^{\left( 0 \right)}  - a_1^{\left(  +  \right)} } \right) \\
 i\dot {a_1}^{\left(  -  \right)}  =  - 2\omega a_1^{\left(  -  \right)}  + \frac{{{\lambda '}_0 }}{2}e^{ - i\delta _ -  t} a_3^{\left(  -  \right)}  \\
 i\dot {a_3}^{\left(  -  \right)}  =  - 2\omega a_3^{\left(  -  \right)}  + \frac{{{\lambda '}_0 }}{2}e^{i\delta _ -  t} \left( {a_1^{\left(  -  \right)}  - a_1^{\left( 0 \right)} } \right) \\
 i\dot {a_1}^{\left(  +  \right)}  = 2\omega a_2^{\left(  +  \right)}  + \frac{{{\lambda }'_0 }}{2}e^{ - i\delta _ -  t} \left( {a_3^{\left(  +  \right)}  - a_3^{\left( 0 \right)} } \right) \\
 i\dot {a_3}^{\left(  +  \right)}  = 2\omega a_3^{\left(  +  \right)}  + \frac{{{\lambda '}_0 }}{2}e^{i\delta _ -  t} a_1^{\left(  +  \right)}  \\
 \end{array} \right..
\end{equation}
Adiabatic elimination yields
\begin{equation}
a_1^{\left(  -  \right)}  \approx 0,\,\,a_3^{\left(  +  \right)}
\approx 0
\end{equation}
and
\begin{equation}
a_1^{\left(  +  \right)}  \approx \frac{{\lambda _0 }}{{4\omega _
-  }}a_3^{\left( 0 \right)} e^{-i\delta _ -  t} ,\,\,a_3^{\left( -
\right)}  \approx  - \frac{{\lambda _0 }}{{4\omega _ -
}}a_1^{\left( 0 \right)} e^{ i\delta _ -  t},
\end{equation}
which in turn amount to the set
\begin{equation}
\left\{ \begin{array}{l}
 i\dot{ a_1}^{\left( 0 \right)}  = \frac{{5{\lambda'} _0^2 }}{{8\omega }}a_1^{\left( 0 \right)}  + \frac{{{\lambda'} _0 }}{2}e^{-i\delta _ -  t} a_3^{\left( 0 \right)}  \\
\\
 i\dot{ a_3}^{\left( 0 \right)}  =  - \frac{{5{\lambda'} _0^2 }}{{8\omega }}a_3^{\left( 0 \right)}  + \frac{{{\lambda'} _0 }}{2}e^{ i\delta _ -  t} a_1^{\left( 0 \right)}  \\
 \end{array} \right..
\end{equation}
Note that this procedure enables us to calculate only the
first-order correction term $\sim {{\lambda '_0 } \mathord{\left/
 {\vphantom {{\lambda '_0 } \omega }} \right.
 \kern-\nulldelimiterspace} \omega }$
 to the RWA since account of the
higher order terms makes the set (22) incompatible.

The substitution $a_1^{\left( 0 \right)}  = \tilde a_1^{\left( 0
\right)} e^{ - i\frac{{5{\lambda'} _0^2 t}}{{8\omega }}}
,\,\,a_3^{\left( 0 \right)} = \tilde a_3^{\left( 0 \right)}
e^{i\frac{{5{\lambda'} _0^2 t}}{{8\omega }}} $
 transforms the set of Eqs. (22) into
\begin{equation}
\left\{ \begin{array}{l}
 i\dot{ \tilde a_1}^{\left( 0 \right)}  = \frac{{{\lambda'} _0 }}{2}e^{ - i\left( {\delta _ -   - \frac{{5{\lambda'} _0^2 }}{{4\omega }}} \right)t} \tilde a_3^{\left( 0 \right)}  \\
\\
 i\dot{ \tilde a_3}^{\left( 0 \right)}  = \frac{{{\lambda'} _0 }}{2}e^{ i\left( {\delta _ -   - \frac{{5{\lambda'} _0^2 }}{{4\omega }}} \right)t} \tilde a_1^{\left( 0 \right)}  \\
 \end{array} \right.
 \end{equation}
that is equivalent to the following second-order differential
equation:
\begin{equation}
\ddot {\tilde a_1}^{\left( 0 \right)}  + i\left( {\delta _ -   -
\frac{{5{\lambda'} _0^2 }}{{4\omega }}} \right)\dot {\tilde
a_1}^{\left( 0 \right)}  + \frac{{{\lambda'} _0^2 }}{4}\tilde
a_1^{\left( 0 \right)}  = 0
\end{equation}
with the initial conditions $\tilde a_1^{\left( 0 \right)} \left(
0 \right) = \frac{{3\lambda _0 }}{{4\sqrt 2 \omega }},\,\,\dot{
\tilde a_1}^{\left( 0 \right)} \left( 0 \right) = -i\frac{{\lambda
_0 }}{{2\sqrt 2 }}$.

Of course, we have to justify the adiabatic approximation used in
Eqs. (20) - (21) by imposing the requirement on the pulse
switching time $\tau _0 $:
\begin{equation}
\omega ^{ - 1}  \ll \tau _0  \ll T.
\end{equation}
In what follows, however, we shall continue to handle the ramp
pulses since the accurate calculation carried out for an
adiabatically switched pulse brings about simple renormalization
of the matrix element $\lambda_0$ conserving the total character
of the ramp-pulsed dynamics \cite{28}. Note that the first of
inequalities (25) ensures the applicability of Eq. (13) at
$t\leq\tau_0$. Besides, we assume that $\tau _0 \ll \left|
{\lambda _0 } \right|^{ - 1} $. This unnecessary but very useful
condition minimizes the influence of the pulse shape on the Rabi
oscillation pattern.

The solution of Eq. (24) is straightforward; transforming it back
to the  coefficients $a_1^{\left( 0 \right)}$ and $a_3^{\left( 0
\right)}$, one has
\begin{equation}
 \begin{array}{l}
 a_1^{\left( 0 \right)}  = \frac{1}{{\sqrt 2 }}e^{ - i\frac{{\delta _ -  t}}{2}} \left[ {\frac{{3\lambda _0 }}{{4\omega }}\cos \left( {2{\Omega '}_ -  t} \right) - i\frac{{\lambda '_0 }}{{4{\Omega '}_ -  }}\sin \left( {2{\Omega '}_ -  t} \right)} \right], \\
 a_3^{\left( 0 \right)}  = \frac{1}{{\sqrt 2 }}e^{i\frac{{\delta _ -  t}}{2}} \left[ {\cos \left( {2{\Omega '}_ -  t} \right) - i\frac{{\delta _ -   - {{{\lambda '}_0^2 } \mathord{\left/
 {\vphantom {{{\lambda '}_0^2 } {2\omega }}} \right.
 \kern-\nulldelimiterspace} {2\omega }}}}{{4\Omega '_ -  }}\sin \left( {2\Omega '_ -  t} \right)} \right], \\
  \end{array}
\end{equation}
where $\Omega '_ -   = {{\sqrt {{\lambda '}_0^2  + {\delta '}_ -
^2 } } \mathord{\left/
 {\vphantom {{\sqrt {{\lambda '}_0^2  + {\delta '}_ - ^2 } } 4}} \right.
 \kern-\nulldelimiterspace} 4},\,\,\,\,\,\delta '_ -   = \delta _ -   - {{5{\lambda '}_0^2 } \mathord{\left/
 {\vphantom {{5{\lambda '}_0^2 } {4\omega }}} \right.
 \kern-\nulldelimiterspace} {4\omega }}$.
 It is easy to verify that these expressions satisfy the set
of Eqs. (19).

 As a result, for the coefficients in the laboratory frame we
obtain
\begin{equation}
\begin{array}{l}
 c_{A0(B0)}  = \frac{1}{2}e^{ - i\varepsilon _0 t} \left[ {f_0^ -  \left( t \right) \pm f_0^ +  \left( t \right)} \right], \\
 f_0^ \pm  \left( t \right) = e^{i\frac{{\delta _ \pm  t}}{2}} \left[ {\cos \left( {2\Omega _ \pm  t} \right) - i\frac{{\delta _ \pm  }}{{4\Omega _ \pm  }}\sin \left( {2\Omega _ \pm  t} \right) - i\frac{{\tilde \lambda ^2 }}{{8\omega \Omega _ \pm  }}e^{ - i2\omega t} \sin \left( {2\Omega _ \pm  t} \right)} \right]; \\
 \end{array}
\end{equation}
\begin{equation}
\begin{array}{l}
 c_{A1(B1)}  = \frac{1}{2}\left[ {e^{ - i\varepsilon _ -  t} f_1^ -  \left( t \right) \pm e^{ - i\varepsilon _ +  t} f_1^ +  \left( t \right)} \right], \\
 f_1^ \pm  \left( t \right) = e^{ - i\frac{{\delta _ -  t}}{2}} \frac{{\tilde \lambda }}{{2\sqrt 2 \Omega _ \pm  }}\left[ { - i\sin \left( {2\Omega _ \pm  t} \right) + \frac{{\Omega _ \pm  }}{\omega }\left( {1 - e^{i2\omega t} } \right)\cos \left( {2\Omega _ \pm  t} \right)} \right]; \\
 \end{array}
\end{equation}
where
\begin{equation}
\Omega _ \pm = \frac{1}{4}\sqrt {2\tilde\lambda ^2  + \delta _ \pm
^2 }
\end{equation}
are the Rabi frequencies and $\tilde\lambda  = \frac{{\lambda _0
}}{{\sqrt 2 }}$, $\varepsilon _ \pm   = \varepsilon _0  + \omega _
\pm $.

The formulae (27) - (29) describe the general type of the coherent
one-electron evolution in the symmetric four-level double-dot
structure driven by the resonant monochromatic pulse. One can
check that the normalization condition
\begin{equation}
\sum\limits_{n = A0,B0,A1,B1} {\left| {c_n \left( t \right)}
\right|} ^2  = 1 + O\left( {{{\lambda _0^2 } \mathord{\left/
 {\vphantom {{\lambda _0^2 } {\omega ^2 }}} \right.
 \kern-\nulldelimiterspace} {\omega ^2 }},{{\delta _ \pm  } \mathord{\left/
 {\vphantom {{\delta _ \pm  } \omega }} \right.
 \kern-\nulldelimiterspace} \omega }} \right)
\end{equation}
is fulfilled with the accuracy adopted through the calculations.

We couldn't reveal any noticeable effect on the dynamics that
would be produced by the small parameters ${{\delta _ \pm  }
\mathord{\left/
 {\vphantom {{\delta _ \pm  } \omega }} \right.
 \kern-\nulldelimiterspace} \omega }$. They appear in the expression for
the Rabi frequencies $\Omega '_ \pm  $, Eq. (26), and may compete
with the terms ${{\delta _ \pm ^2 } \mathord{\left/
 {\vphantom {{\delta _ \pm ^2 } {\lambda _0^2 }}} \right.
 \kern-\nulldelimiterspace} {\lambda _0^2 }}$
 if ${{\left| {\lambda _0 } \right|} \mathord{\left/
 {\vphantom {{\left| {\lambda _0 } \right|} \omega }} \right.
 \kern-\nulldelimiterspace} \omega } \ge \left| {{{\delta _ \pm  } \mathord{\left/
 {\vphantom {{\delta _ \pm  } \lambda }} \right.
 \kern-\nulldelimiterspace} \lambda }_0 } \right|$.
 Since we are not interested in detailed consideration of the
system dynamics in that range of parameters we have neglected the
terms $ \sim {{\delta _ \pm  } \mathord{\left/
 {\vphantom {{\delta _ \pm  } \omega }} \right.
 \kern-\nulldelimiterspace} \omega }$ in Eqs. (27) - (29).
 Note that the terms ${{\lambda _0 \delta _ \pm  } \mathord{\left/
 {\vphantom {{\lambda _0 \delta _ \pm  } {\omega V }}} \right.
 \kern-\nulldelimiterspace} {\omega V }},\,{{\lambda _0 \delta _ \pm  } \mathord{\left/
 {\vphantom {{\lambda _0 \delta _ \pm  } {\omega ^2 }}} \right.
 \kern-\nulldelimiterspace} {\omega ^2 }}$ which also contain this small parameter may be comparable to the
terms of the order of ${{\lambda _0^2 } \mathord{\left/
 {\vphantom {{\lambda _0^2 } {\omega ^2 }}} \right.
 \kern-\nulldelimiterspace} {\omega ^2 }}$ that have been omitted in the solution, and hence must be
omitted as well.

\vskip 6mm

\centerline{\bf III. QUANTUM DYNAMICS IN THE STRONG AND WEAK }
\centerline{\bf TUNNELING REGIMES.}

In this Section we analyze the results obtained above for various
choices of the pulse and structure parameters. From the general
formulae (27) - (29), we derive the expressions for the
probability amplitudes corresponding to the situations where the
characteristic tunnelling energy $V$ is either much greater or
much less than the matrix element of optical dipole transition
$\left|\lambda_0 \right|$. Besides, we investigate also the
electron dynamics in the strongly detuned Raman-like regime.

\centerline{\bf A. The three-level quantum dynamics.}

 If an electron being excited in one of the QDs tunnels
into the other QD during a time much shorter than the Rabi
oscillation period, one can speak about the simultaneous
electronic excitation in both of QDs. In other words, one of the
hybridized states, i.e. ${{\left( {\left| {A1} \right\rangle  +
\left| {B1} \right\rangle } \right)} \mathord{\left/
 {\vphantom {{\left( {\left| {A1} \right\rangle  + \left| {B1} \right\rangle } \right)} {\sqrt 2 }}} \right.
 \kern-\nulldelimiterspace} {\sqrt 2 }}$ or ${{\left( {\left| {A1} \right\rangle  - \left| {B1} \right\rangle } \right)} \mathord{\left/
 {\vphantom {{\left( {\left| {A1} \right\rangle  - \left| {B1} \right\rangle } \right)} {\sqrt 2 }}} \right.
 \kern-\nulldelimiterspace} {\sqrt 2 }}$ (equally-weighted in each of the QDs), is excited. Both of
 these states are the eigenstates of the stationary Hamiltonian $H_{0}$, Eq. (1).
It seems then preferable to expand the state vector $\left| {\Psi
\left( t \right)} \right\rangle $ over the eigenstates of $H_{0}$
instead than over the states of isolated QD basis. Doing so and
using the resonant approximation we may consider only one of
hybridized states that is formally equivalent to the setting $V\gg
\left|\lambda_0 \right|$. The quantum dynamics of our system thus
coincides with that of the three-level nanostructure.
 This situation was studied in the works \cite{18} - \cite{24} without, however,
paying enough attention to the mathematical proof of that
proposal. In what follows we shall consider this case in detail
and show to what extent the electron dynamics may correspond to
the scheme just sketched.

The condition of the strong tunnel coupling between the excited
orbital states of the QDs, as compared to the optical dipole
coupling between the ground and excited orbital states of the
single QD, is
\begin{equation}
2V = \left| {\delta _ +   - \delta _ -  } \right| \gg \left|
\lambda_0 \right|.
\end{equation}
We are interested in the resonant electron-pulse interaction, when
the pulse frequency $\omega$ matches one of the resonant
frequencies $\omega _ {\pm}$ and is strongly detuned from the
other one. For definiteness, let the pulse frequency $\omega$ to
be close to the frequency $\omega _ {-}$ so that  $\left| {\delta
_ -  } \right| \ll \left| {\delta _ +  } \right|$ and, as it
follows from Eq. (31),  $\left| {\delta _ +  } \right| \gg \left|
\lambda_0  \right|$. This choice of the pulse frequency
corresponds to electron transition from $\left| {A(B)0}
\right\rangle$ to ${{\left( {\left| {A1} \right\rangle  + \left|
{B1} \right\rangle } \right)} \mathord{\left/
 {\vphantom {{\left( {\left| {A1} \right\rangle  + \left| {B1} \right\rangle } \right)} {\sqrt 2 }}} \right.
 \kern-\nulldelimiterspace} {\sqrt 2 }}$. Obviously, there are two different situations
concerning the mutual relation between the pulse detuning $\delta
_ - $ and the value of $\left| \lambda_0 \right|$,
 i.e. $\left| {\delta _ -  } \right| \ll \left| \lambda_0  \right|$ and
$\left| {\delta _ -  } \right| \gg \left| \lambda_0  \right|$. The
first inequality corresponds to the resonant electron-pulse
interaction, whereas the second one describes the off-resonant
Raman-like coupling (see Sec. III C).

Since here we consider the resonant case, all of the conditions
imposed on the system parameters may be summarized in the
following inequalities:
\begin{equation}
\left| {\delta _ -  } \right| \ll \left| \lambda_0  \right| \ll
\left| {\delta _ +  } \right|,
\end{equation}
where $\delta _ +   \approx  - 2V$ and $\omega _ -   = \omega _ +
- 2V $.

Making use of Eqs. (27) - (29) and taking into account that
\begin{equation}
\cos \left( {2\Omega_{\pm} t} \right) - i\frac{\delta_{\pm}
}{{4\Omega_{\pm} }}\sin \left( {2\Omega_{\pm} t} \right) \approx
e^{ - i\frac{\delta_{\pm} }{2}\left( {1 + \frac{{\tilde\lambda ^2
}}{{\delta_{\pm} ^2 }}} \right)t}  + i{\mathop{\rm sgn}} \left(
\delta_{\pm} \right)\frac{{\tilde\lambda ^2 }}{{\delta_{\pm} ^2
}}\sin \left( {\frac{{\left| \delta_{\pm}  \right|}}{2}t} \right)
\end{equation}
at $\left| \lambda_0  \right| \ll \left| \delta_{\pm} \right|$,
let us rewrite the Eq. (3) in the form
\begin{equation}
\begin{array}{l}
 \left| \Psi  \right\rangle  = \frac{1}{2}e^{ - i\varepsilon _0 t} \left[ {1 + f_0^ -  \left( t \right) - i\frac{{\tilde \lambda ^2 }}{{4V ^2 }}e^{ - iVt} \sin \left(V t \right)} \right]\left| {A0} \right\rangle  +  \\
 \,\,\,\,\,\,\,\,\, + \frac{1}{2}e^{ - i\varepsilon _0 t} \left[ { - 1 + f_0^ -  \left( t \right) + i\frac{{\tilde \lambda ^2 }}{{4V ^2 }}e^{ - iVt} \sin \left(V t \right)} \right]\left| {B0} \right\rangle  +  \\
  + \frac{1}{{\sqrt 2 }}e^{ - i\varepsilon _ -  t} f_1^ -  \left( t \right)\frac{{\left| {A1} \right\rangle  + \left| {B1} \right\rangle }}{{\sqrt 2 }}\, - i\frac{{\tilde \lambda }}{2V }e^{ - i\left( {\varepsilon _ +   - V} \right)t} \sin \left(V t \right)\frac{{\left| {A1} \right\rangle  - \left| {B1} \right\rangle }}{{\sqrt 2 }}. \\
 \end{array}
\end{equation}
It is easy to calculate from Eq. (34) the probability of the state
inversion $p_{B0} \left( t \right) = \left| {\left\langle {{B0}}
 \mathrel{\left | {\vphantom {{B0} \Psi \left( t \right)}}
 \right. \kern-\nulldelimiterspace}
 {\Psi \left( t \right)} \right\rangle } \right|^2 $ after the pulse of the
 duration $T_\pi   = {\pi  \mathord{\left/
 {\vphantom {\pi  {\lambda_0 }}} \right.
 \kern-\nulldelimiterspace} {\lambda_0 }}$ (the so-called
 $\pi$-pulse) is off:
\begin{equation}
p_{B0} \left( {T_\pi  } \right) = 1 - \frac{{\delta _ - ^2 \pi ^2
}}{{64\Omega _ - ^2 }} - \frac{{\tilde \lambda ^2 }}{{4V ^2 }}\sin
^2 \left( {\frac{{V \pi }}{{2\Omega _ -  }}} \right)
\end{equation}

The first two terms in Eq. (35) correspond to the results of Ref.
\cite{18} where the off-resonant electron transitions to the state
${{\left( {\left| {A1} \right\rangle  - \left| {B1} \right\rangle
} \right)} \mathord{\left/
 {\vphantom {{\left( {\left| {A1} \right\rangle  - \left| {B1} \right\rangle } \right)} {\sqrt 2 }}} \right.
 \kern-\nulldelimiterspace} {\sqrt 2 }}$ were completely neglected.
 The third term arises due to account for such transitions. Their
 contribution to the electron state evolution (34) is proportional to the small parameter $\frac{\tilde\lambda }{V
}\ll 1$ and results in the relative phase and amplitude shifts
between the coefficients $c _ {A1}$ and $c _ {B1}$ that indicate
on the finite tunnelling time $\tau _{tunn} \sim {1
\mathord{\left/
 {\vphantom {1 2V }} \right.
 \kern-\nulldelimiterspace} V }$
 between the QDs. Moreover, the oscillations at the frequency
$2\omega$ (so-called Bloch-Siegert oscillations) affect, to some
extent, the ideal three-level oscillation picture.

The results obtained show that the three-level scheme can be used
for the description of electron dynamics if the conditions
$\frac{\tilde\lambda }{V } \ll 1,\,\,\,\frac{\tilde\lambda
}{\omega } \ll 1$ are satisfied. The errors introduced due to the
presence of a nearby forth level are of the order of
$\frac{\tilde\lambda }{V }$.

\centerline{\bf B. The two-level quantum dynamics.}

Next we study the opposite case of small $V$ when a substantial
spreading of an excited electron between the QDs occurs after many
Rabi oscillations in single QD have completed. Such an electron
dynamics is realized in the double-dot structure where the tunnel
coupling between the excited states of the QDs is rather small as
compared to the electron-pulse coupling:
\begin{equation}
2V = \left| {\delta _ +   - \delta _ -  } \right| \ll \left|
\lambda_0  \right|.
\end{equation}
The condition (36) may be satisfied for two different pulse
designs, i.e. for both $\left| \lambda_0  \right| \ll \left|
{\delta _ -  } \right|,\left| {\delta _ +  } \right|$ and $\left|
{\delta _ -  } \right|,\left| {\delta _ +  } \right| \ll \left|
\lambda_0 \right|$. The first inequality corresponds to the
off-resonant single QD excitation whereas the second one
characterizes the resonant two-level Rabi oscillations in the same
QD. The off-resonant case does not reveal significant two-level
dynamics since an electron stays predominantly localized in the
state ${\left| {A0} \right\rangle }$ (the population of the state
${\left| {A1} \right\rangle }$ is of the order of ${{\lambda_0 ^2
} \mathord{\left/
 {\vphantom {{\lambda ^2 } {\delta _ \pm ^2 }}} \right.
 \kern-\nulldelimiterspace} {\delta _ \pm ^2 }} \ll 1$).
 We shall focus our attention on the resonant transition for which the
conditions
\begin{equation}
V,\left| {\delta _ -  } \right|,\left| {\delta _ +  } \right| \ll
\left| \lambda_0  \right|
\end{equation}
are satisfied.

In this case an electron oscillates between the ground and excited
states of the QD A for the pulse durations $T \ll {1
\mathord{\left/
 {\vphantom {1 V }} \right.
 \kern-\nulldelimiterspace} V }$. To prove this statement we make use of the approximations
$\Omega _ +   - \Omega _ -   \approx 2V \frac{{\delta _ -
}}{{\left| \tilde\lambda  \right|}}$ and $e^{ - iVt} \approx 1 -
iVt$ that holds for the time domain $t \ll {1 \mathord{\left/
 {\vphantom {1 V }} \right.
 \kern-\nulldelimiterspace} V }$. Inserting them into the Eqs. (27)-(29) and retaining in the time
dependencies the terms up to the first order in $V  t$, we have
the following expression for the state vector:
\begin{equation}
\begin{array}{l}
 \left| \Psi  \right\rangle  = e^{ - i\varepsilon {}_0t} \left\{ {\left[ {f_0^ -  \left( t \right) + i\frac{{t \cdot V }}{2}e^{i\frac{{\delta _ -  t}}{2}} \cos \left( {2\Omega _ -  t} \right)} \right]\left| {A0} \right\rangle } \right. + \left. {i\frac{{t \cdot V }}{2}e^{i\frac{{\delta _ -  t}}{2}} \cos \left( {2\Omega _ -  t} \right)\left| {B0} \right\rangle } \right\} +  \\
 \, + e^{ - i\varepsilon _ -  t} \left\{ {\left[ {f_1^ -  \left( t \right) + \frac{{\tilde \lambda }}{{2\sqrt 2 \Omega _ -  }}\frac{{t \cdot V }}{2}e^{ - i\frac{{\delta _ -  t}}{2}} \sin \left( {2\Omega _ -  t} \right)} \right]\left| {A1} \right\rangle } \right. - \left. {\frac{{\tilde \lambda }}{{2\sqrt 2 \Omega _ -  }}\frac{{t \cdot V }}{2}e^{ - i\frac{{\delta _ -  t}}{2}} \sin \left( {2\Omega _ -  t} \right)\left| {B1} \right\rangle } \right\} \\
 \end{array}
\end{equation}
that demonstrates the two-level electron evolution in the QD A is
slightly distorted by the excitations in the QD B. Setting in Eq.
(38) $V = 0, \delta_{-} =0$ we find the state-vector evolution for
the two-level system being in the exact resonance with the
external pulse
\begin{equation}
\begin{array}{l}
 \left| \Psi  \right\rangle  = e^{ - i\varepsilon _0 t} \left[ {\cos \left( {\lambda_0 t/2} \right) - i\frac{\tilde\lambda }{{4\omega }} e^{ - i2\omega t} \sin \left( {\lambda_0 t/2} \right)} \right]\left| {A0} \right\rangle  +  \\
 \,\,\,\,\,\,\,\, + e^{ - i\varepsilon _ -  t} {\left[ { - i\sin \left( {\lambda_0 t/2} \right) + \frac{\tilde\lambda }{{4\omega }}\left( {1 - e^{i2\omega t} } \right)\cos \left( {\lambda_0 t/2} \right)} \right]\left| {A1} \right\rangle } .\\
 \end{array}
\end{equation}

If one adopts the scheme where the quantum information is encoded
into the ground and first excited electron (or exciton) states of
the single QD (\cite{6}, \cite{8}, \cite{29}), the population
transfer into neighboring QD should be considered as the
information leakage from the computational subspace and the
corresponding error probability may be evaluated with the help of
Eq. (38). The computational error introduced by the Bloch-Siegert
oscillations alone is deduced from Eq. (39).

\centerline{\bf C. The electron excitation driven by the strongly
detuned pulse. }

Finally, we shall examine the case of the Raman-like off-resonant
excitation of an electron in the double-dot structure ($\left|
\lambda_0  \right| \ll \left| {\delta _ {\pm}  } \right|$). This
mechanism of the optical quantum-state engineering is currently
under extensive investigations because of important properties
that distinguish it from the other optical schemes (see, e.g.,
\cite{30}). First, the excited (auxiliary) states are populated
only virtually that allows one to localize the electron population
 almost completely in the ground-state subspace
$\left\{ {\left| {A0} \right\rangle ,\left| {B0} \right\rangle }
\right\}$. Provided that the states $\left| {A0} \right\rangle$
and $\left| {B0} \right\rangle$ constitute the qubit basis one can
therefore operate with the quantum information trapped in the
logical subspace for any time. This, in its turn, simplifies the
state evolution design and prevents the qubit from the decoherence
induced by spontaneous photon emission from the excited levels.
Second, the population transfer realized via the off-resonant
excitations is quite robust against the pulse imperfections such
as the uncontrollable detunings and the timing errors. The quantum
optics provides one with wide class of schemes specially developed
for those purposes. Quite recently several attempts have been made
to adopt those schemes for the solid-state objects possessing of
the atomic-like spectrum, e.g. the QDs \cite{21}, the QDs combined
with cavity QEDs \cite{7}, the rf-SQUIDs \cite{31}.

Some features of the qubit state evolution based on one-electron
quantum dynamics in the symmetric double-dot structure driven by
the strongly detuned pulse have been outlined in the work
\cite{21}. Here we consider this effect as the particular case of
the four-level double-dot dynamics studied in Sec. II. Choosing
the system parameters to satisfy the inequalities
\begin{equation}
\left| \lambda_0  \right| \ll \left| {\delta _ -  } \right|,\left|
{\delta _ +  } \right| \ll \omega ,\,\,\,\left| \lambda_0  \right|
\ll V,
\end{equation}
we get from the general formulae (27)-(29) the following
expression for the state vector
\begin{equation}
\left| \Psi \right\rangle \approx e^{ - i\left( {\varepsilon _0  +
\frac{{\delta _ -   + \delta _ +  }}{{4\delta _ -  \delta _ +
}}\tilde\lambda ^2 } \right)t} \left[ {\cos \left( {\frac{V
}{{2\delta _ -  \delta _ + }}\tilde\lambda ^2 t} \right)\left|
{A0} \right\rangle  + i\sin \left( {\frac{V }{{2\delta _ - \delta
_ +  } }\tilde\lambda ^2 t} \right)\left| {B0} \right\rangle }
\right].
\end{equation}
The Eq. (41) describes the two-level Rabi oscillations at the
frequency $\Omega _\delta   = {{V \tilde\lambda ^2 }
\mathord{\left/
 {\vphantom {{V \tilde\lambda ^2 } {2\delta _ -  \delta _ +  }}} \right.
 \kern-\nulldelimiterspace} {2\delta _ -  \delta _ +  }}$
 similar to those induced by the electrostatic field
(see Eq. (14)). The Rabi frequencies of these processes are very
different from each other, viz. ${{\Omega _0 } \mathord{\left/
 {\vphantom {{\Omega _0 } \Omega _ \delta }} \right.
 \kern-\nulldelimiterspace} \Omega _ \delta } \sim {{\left| {\delta _ -  \delta _ +  } \right|} \mathord{\left/
 {\vphantom {{\left| {\delta _ -  \delta _ +  } \right|} {\omega _{10}^2 }}} \right.
 \kern-\nulldelimiterspace} {\omega _{10}^2 }} \ll 1$. This makes the optically-driven
oscillations more preferable for a qubit-state engineering due to
their higher speed as compared with that of the electrostatic
driving. As it follows from Eqs. (14) and (41) the quantum
dynamics in both cases is frozen if ${{V} \mathord{\left/
 {\vphantom {{V} {\left| \lambda_0  \right|}}} \right.
 \kern-\nulldelimiterspace} {\left| \lambda_0  \right|}}$ approaches zero. This effect may be
explained in terms of the destructive interference between the
probability amplitudes of the hybridized states represented by the
symmetric and antisymmetric superpositions of the excited states
of isolated QDs. Since those states become nearly-degenerate with
the decrease of the parameter ${{V} \mathord{\left/
 {\vphantom {{V} {\left| \lambda_0  \right|}}} \right.
 \kern-\nulldelimiterspace} {\left| \lambda_0  \right|}}$, their probability amplitudes sum up
in QD A whereas they cancel one another in QD B. Despite of these
states are empty during the population transfer, the importance of
their assistance to the process becomes more clear from this
analysis.

It is worth to note that the system evolution described by Eq.
(41) cannot result in an arbitrary rotation of the qubit-state
vector on the Bloch sphere since it contains only one
time-dependent parameter $\theta  = \Omega_ \delta t$
corresponding to the polar angle (the azimuthal angle is fixed and
equals to ${\pi \mathord{\left/
 {\vphantom {\pi  2}} \right.
 \kern-\nulldelimiterspace} 2}$). To overcome this obstacle one should break the
symmetry of the structure and use at least two driving pulses with
different parameters to implement the desired rotation \cite{20}.

\vskip 7mm

\centerline{\bf IV. NUMERICS }

To illustrate the analytical results of Sec. II and Sec. III we
have performed the numerical simulations of electron dynamics in
our structure. The Eq. (5) with $\lambda$ defined by the Eq. (16)
was integrated within the time interval $0 \le t \le 3T$ (where
$T=\pi/\lambda_0$) for ${{\left| \lambda_0  \right|}
\mathord{\left/
 {\vphantom {{\left| \lambda_0  \right|} \omega }} \right.
 \kern-\nulldelimiterspace} \omega } = 10^{ - 3} $ and $\delta _ -   = 0$.
This choice of the pulse-structure parameters corresponds to that
usually realized in the QD systems where $\omega _{10} \sim 10^{ -
2} $ eV and $\left| {\lambda _0 } \right| \sim 10^{ - 5} $ eV for
the pulse strength $E_0 \sim 1 \div 10$ V/cm. Since we are
interested in demonstration of the transition from the three-level
double-dot scheme to the two-level single-dot scheme, the ratio
${{V} \mathord{\left/
 {\vphantom {{V} {\left| \lambda_0  \right|}}} \right.
 \kern-\nulldelimiterspace} {\left| \lambda_0  \right|}}$ was varied from 0.01 to 10.

The numerical plots showing the time dependencies of the
populations $p_n  = \left| {c_n } \right|^2 ,\,\,\,n =
A0,B0,A1,B1$ are presented in the Figs. 2 (a)-(d) for ${{V}
\mathord{\left/
 {\vphantom {{V} {\left| \lambda_0  \right|}}} \right.
 \kern-\nulldelimiterspace} {\left| \lambda_0  \right|}} = 5;1;0.3;0.05$, respectively.
For large but finite values of ${{V} \mathord{\left/
 {\vphantom {{V} {\left| \lambda_0  \right|}}} \right.
 \kern-\nulldelimiterspace} {\left| \lambda_0  \right|}}$ the three-level Rabi oscillations picture
involving the states ${\left| {A0} \right\rangle ,\left| {B0}
\right\rangle ,{{\left( {\left| {A1} \right\rangle  + \left| {B1}
\right\rangle } \right)} \mathord{\left/
 {\vphantom {{\left( {\left| {A1} \right\rangle  + \left| {B1} \right\rangle } \right)} {\sqrt 2 }}} \right.
 \kern-\nulldelimiterspace} {\sqrt 2 }}}$ becomes non-ideal due to the excitation of the state
${{{\left( {\left| {A1} \right\rangle  - \left| {B1} \right\rangle
} \right)} \mathord{\left/
 {\vphantom {{\left( {\left| {A1} \right\rangle  - \left| {B1} \right\rangle } \right)} {\sqrt 2 }}} \right.
 \kern-\nulldelimiterspace} {\sqrt 2 }}}$, see Eq. (34). This effect is clearer seen in the
representation of isolated QD basis since the phase and amplitude
shifts between $p_{A1} $ and $p_{B1} $ provide us with the measure
 characterizing the difference between the electron populations of
the excited levels in the QDs A and B. We see that for ${{V}
\mathord{\left/
 {\vphantom {{V} {\left| \lambda_0  \right|}}} \right.
 \kern-\nulldelimiterspace} {\left| \lambda_0  \right|}} = 5$ (Fig. 2(a)) the results of Sec. III A may be still
applied while for ${{V} \mathord{\left/
 {\vphantom {{V} {\left| \lambda_0  \right|}}} \right.
 \kern-\nulldelimiterspace} {\left| \lambda_0  \right|}}=1$
 (Fig. 2(b)) the regular oscillation pattern is
destroyed when $t \ge 3T$ and the identification of excitation
process is hardly possible. Figure 2(b) clearly demonstrates the
strong modulations of the optically-induced Rabi oscillations
caused by the electron tunnelling if $V \approx \left| \lambda_0
\right|$. Further reduction of the ratio ${{V} \mathord{\left/
 {\vphantom {{V} {\left| \lambda_0  \right|}}} \right.
 \kern-\nulldelimiterspace} {\left| \lambda_0  \right|}}$ amounts to the qualitative changes
in the population dynamics. Figure 2(c) indicates the importance
of the single-QD processes even for ${{V} \mathord{\left/
 {\vphantom {{V} {\left| \lambda_0  \right|}}} \right.
 \kern-\nulldelimiterspace} {\left| \lambda_0  \right|}} = 0.3$. When
${{V} \mathord{\left/
 {\vphantom {{V} {\left| \lambda_0  \right|}}} \right.
 \kern-\nulldelimiterspace} {\left| \lambda_0  \right|}}=0.05$ (Fig.
 2(d)) we observe several almost ideal two-level Rabi oscillations
 in the QD A slightly modulated by residual dynamics in the QD B.
 The total reorganization of oscillation
pattern that marks the transition from one excitation scheme to
another occurs for ${{V} \mathord{\left/
 {\vphantom {{V} {\left| \lambda_0  \right|}}} \right.
 \kern-\nulldelimiterspace} {\left| \lambda_0  \right|}} \sim 0.01$.
The reduction of the Rabi frequency by a factor of 2 and the
depopulation of the states belonging to the QD B are clearly seen
from the numerical plots that confirms the results obtained above.

\vskip 5mm

\centerline{\bf V. CONCLUSIONS}

There are a lot of proposals for the qubit design that use the
basic quantum properties of low-dimensional objects to encode, to
process, and to store the quantum information. The existence of
purely theoretical frameworks is of great importance since they
allow us to capture the principal aspects of idealized evolution
of the system under consideration and then to examine it further
at the more profound level. An exact solution describing the qubit
dynamics is often readily achieved due to the simplified structure
of the model. It is therefore desirable to look for the model that
would include the main features characterizing the qubit and, at
the same time, enable the analytical treatment of the dynamical
problem.

In this paper we have studied in detail the one-electron
double-dot structure proposed as the candidate for a qubit
implementation \cite{18}. The quantum operations in the structure
may be realized by applying the resonant electromagnetic pulse
driving an electron between the QDs. Within this scheme we have
generalized the results recently obtained for several different
pulse-structure setups \cite{18}, \cite{21}, \cite{28} and have
pointed on some delicate aspects concerning their applicability
which, to our knowledge, had never been clarified before. As we
have shown, the efficiency of one or another scheme is conditioned
by the value of the ratio between the matrix element of electron
tunnelling and the matrix element of optical dipole transition.
The mathematical model of the one-electron excitation process has
permitted to study the coherent evolution of the system beyond the
rotating-wave approximation. The numerical results have confirmed
those obtained analytically for the parameter choices
corresponding to the three- and two-level dynamics.

The results presented in this paper may be also applied to the
two-electron symmetric double-dot structure \cite{32} and to the
other systems possessing the same spectral properties, say, to the
superconducting devices \cite{31,33}. Besides we suppose that the
effect of the structure asymmetry on the electron dynamics
\cite{22} may be treated in the same way.

\vskip 4mm

\centerline{\bf ACKNOWLEDGMENTS}

Author thanks L.A. Openov for the stimulating discussions and
useful comments.

\newpage

\includegraphics[width=\hsize]{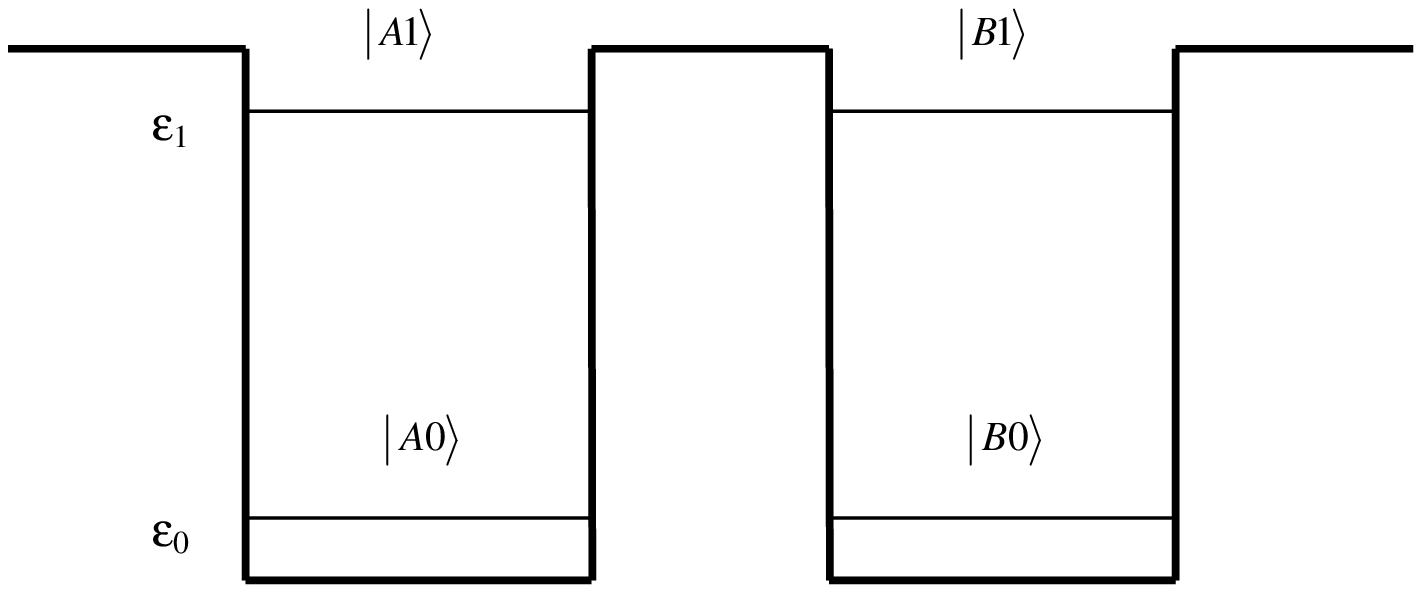}

\vskip 6mm

Fig. 1. Schematics of the states for a single electron confined in
the double-dot structure.
\newpage

\vskip 100mm

Fig. 2. The electron populations of the ground states $\left| {A0}
\right\rangle$ (solid line), $\left| {B0} \right\rangle$ (solid
dashed line) and the excited states $\left| {A1} \right\rangle $
(thin line), $\left| {B1} \right\rangle$ (thin dashed line) of the
symmetric double-dot structure versus the pulse duration $T$ (in
units of $\lambda_0^{-1}$) for the pulse-structure parameter
choice $\delta_- =0$, ${{\left| \lambda_0  \right|}
\mathord{\left/
 {\vphantom {{\left| \lambda_0  \right|} \omega }} \right.
 \kern-\nulldelimiterspace} \omega } = 10^{ - 3} $ and (a) ${{V}
\mathord{\left/
 {\vphantom {{V} {\left| \lambda_0  \right|}}} \right.
 \kern-\nulldelimiterspace} {\left| \lambda_0  \right|}} = 5$, (b) ${{V}
\mathord{\left/
 {\vphantom {{V} {\left| \lambda_0  \right|}}} \right.
 \kern-\nulldelimiterspace} {\left| \lambda_0  \right|}} = 1$, (c) ${{V}
\mathord{\left/
 {\vphantom {{V} {\left| \lambda_0  \right|}}} \right.
 \kern-\nulldelimiterspace} {\left| \lambda_0  \right|}} = 0.3$, (d)
 ${{V}
\mathord{\left/
 {\vphantom {{V} {\left| \lambda_0  \right|}}} \right.
 \kern-\nulldelimiterspace} {\left| \lambda_0  \right|}} = 0.05$.

\newpage

\includegraphics[width=\hsize]{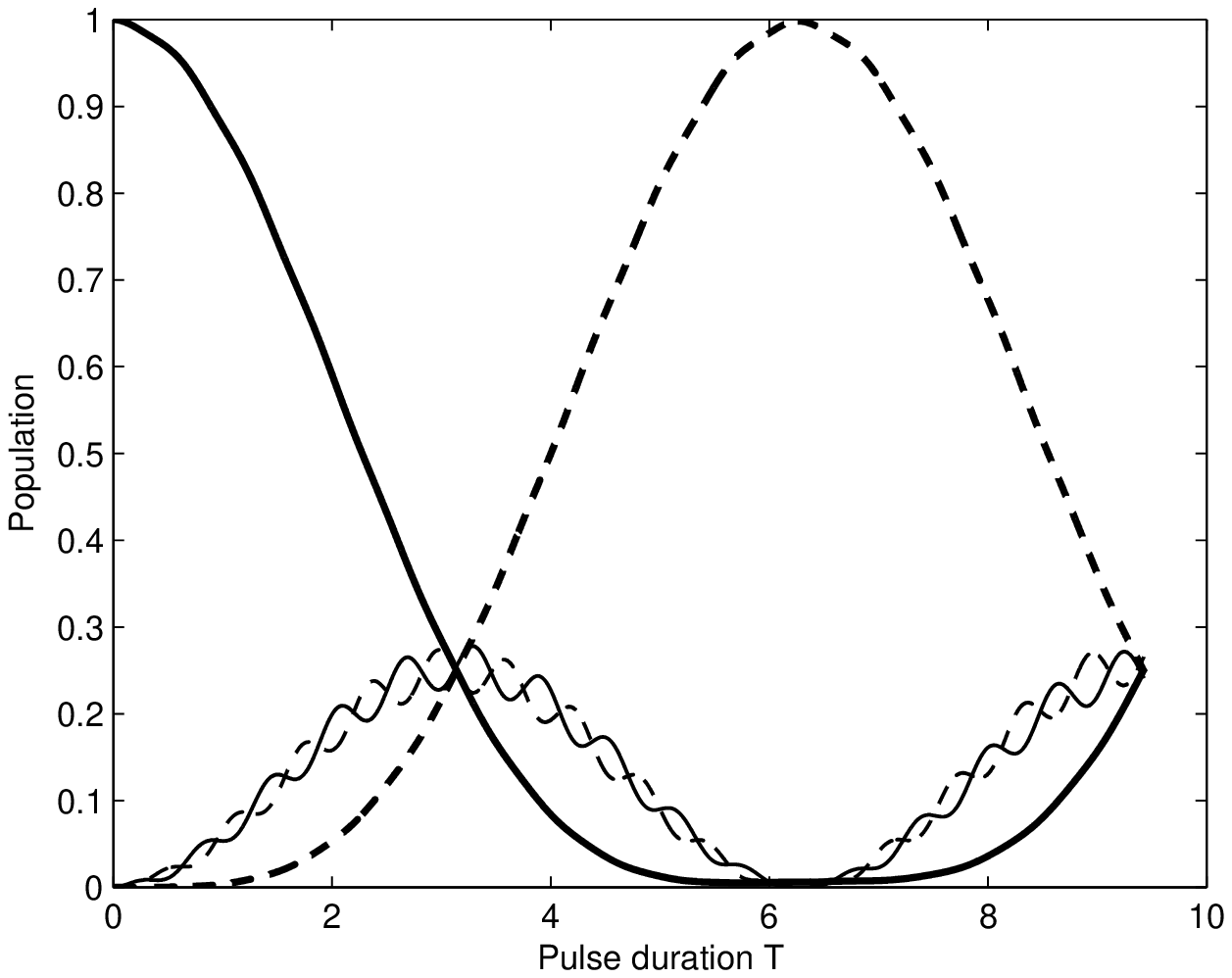}

\vskip 6mm

\centerline{Fig. 2(a)}

\newpage

\includegraphics[width=\hsize]{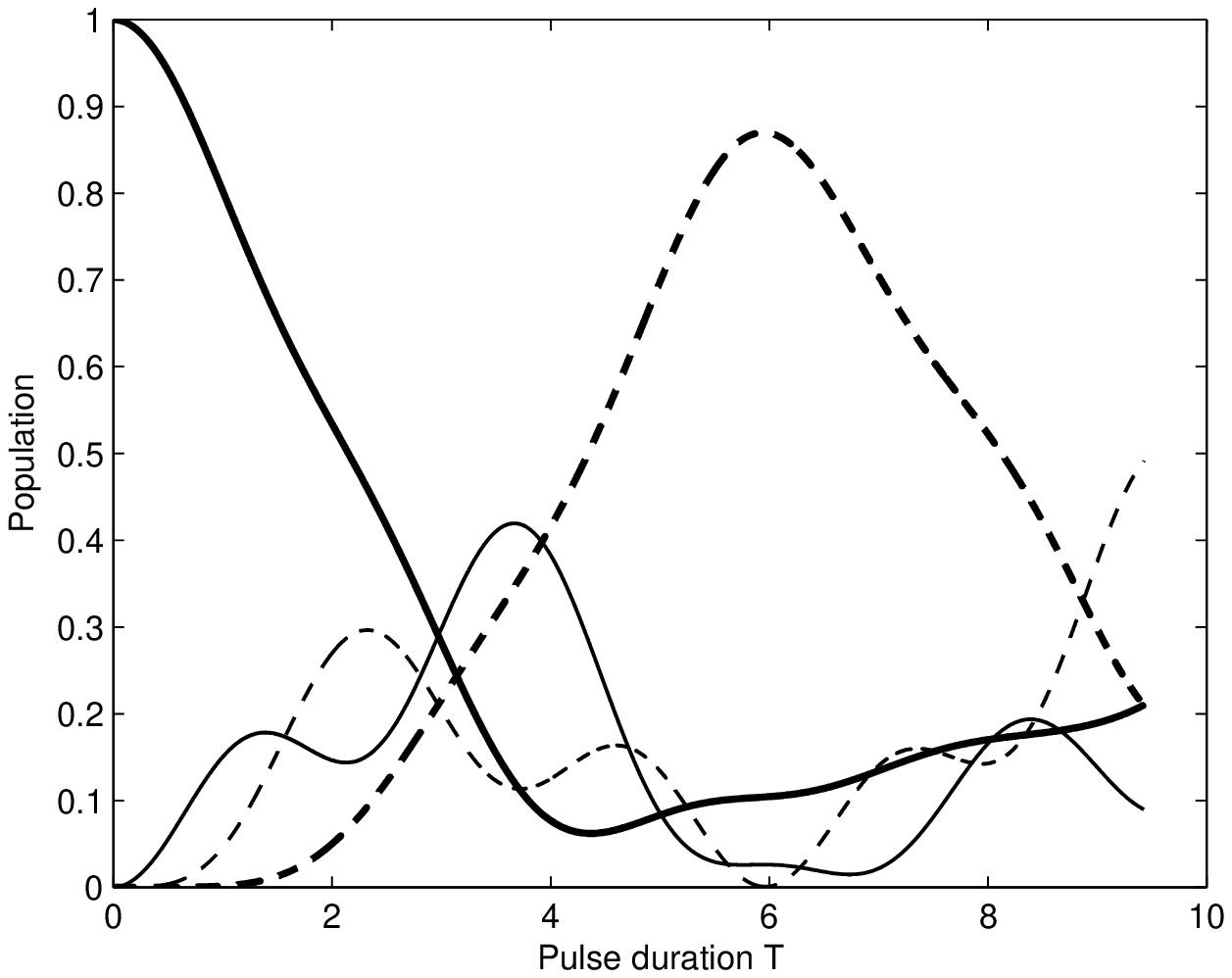}

\vskip 6mm

\centerline{Fig. 2(b)}

\newpage

\includegraphics[width=\hsize]{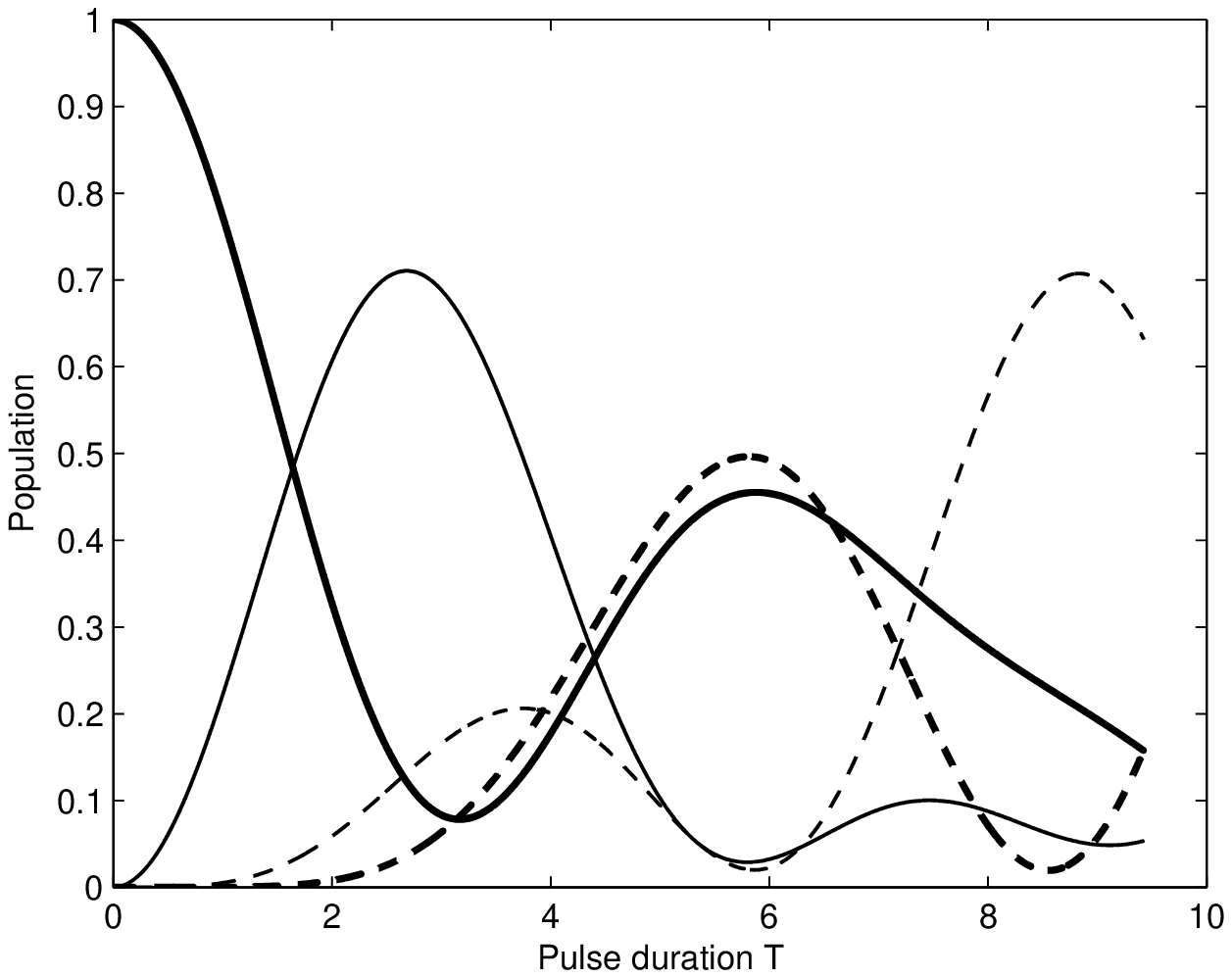}

\vskip 6mm

\centerline{Fig. 2(c)}

\newpage

\includegraphics[width=\hsize]{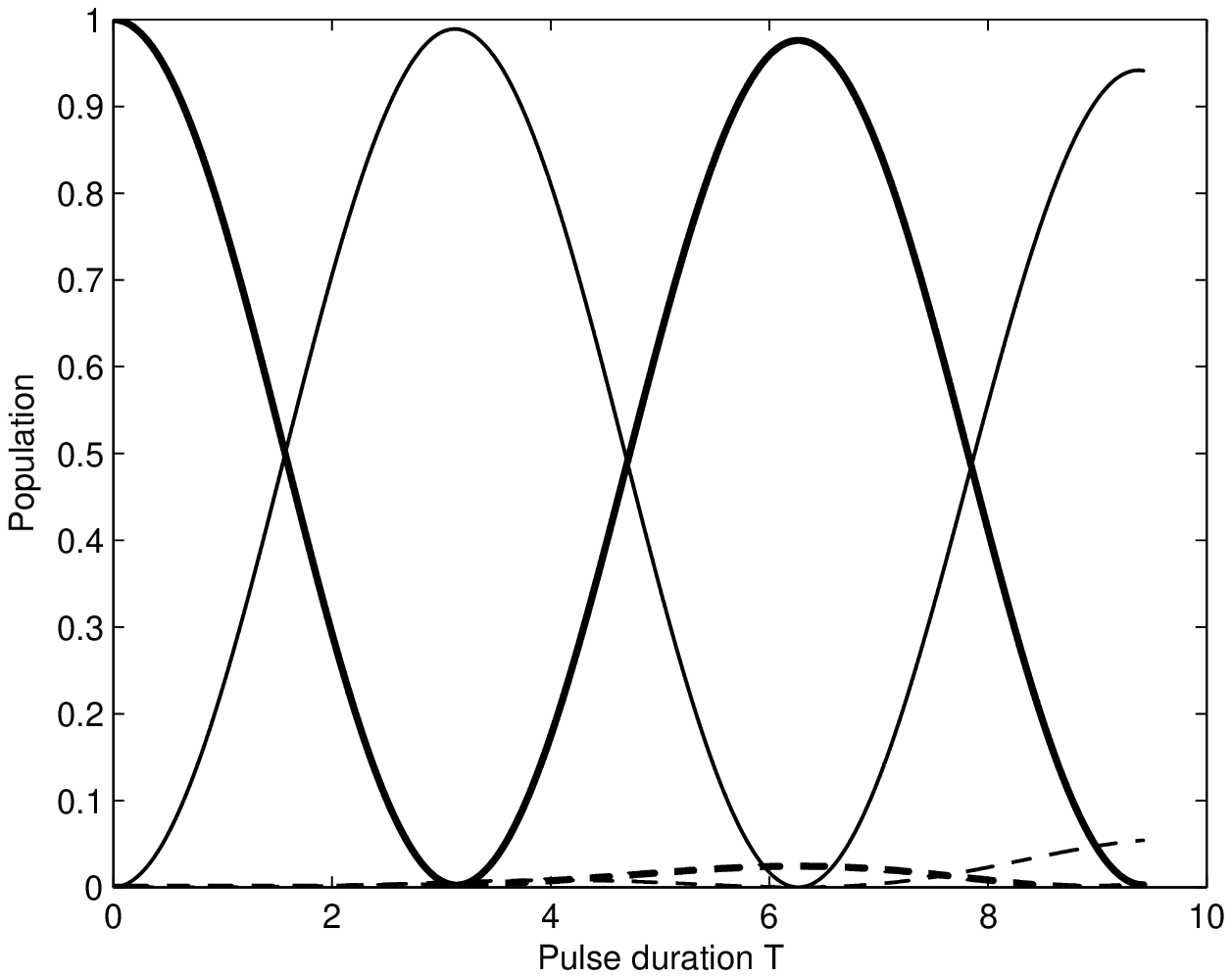}

\vskip 6mm

\centerline{Fig. 2(d)}

\end{document}